\documentclass{iaus}
\usepackage{graphicx}

  \checkfont{eurm10}
  \iffontfound
    \IfFileExists{upmath.sty}
      {\typeout{^^JFound AMS Euler Roman fonts on the system,
                   using the 'upmath' package.^^J}%
       \usepackage{upmath}}
      {\typeout{^^JFound AMS Euler Roman fonts on the system, but you
                   dont seem to have the}%
       \typeout{'upmath' package installed. iaus.cls can take advantage
                 of these fonts,^^Jif you use 'upmath' package.^^J}%
      }
  \else
  \fi

  \checkfont{msam10}
  \iffontfound
    \IfFileExists{amssymb.sty}
      {\typeout{^^JFound AMS Symbol fonts on the system, using the
                'amssymb' package.^^J}%
       \usepackage{amssymb}%

      }{}
  \fi

  \IfFileExists{amsbsy.sty}
    {\typeout{^^JFound the 'amsbsy' package on the system, using it.^^J}%
     \usepackage{amsbsy}}
    {}

\title[Outskirts of Galaxy Clusters: intense life in the suburbs]
      {Dynamical Masses of RCS Galaxy Clusters}

\author[K.~Blindert {\it et al.\/}]%
{Kris Blindert$^1$,
H.~K.~C.~Yee$^1$,
M.~D.~Gladders$^2$,
\and E.~Ellingson$^3$}

\affiliation{
$^1$Department of Astronomy \& Astrophysics, University of Toronto, 60 St.~George St., Toronto, Ontario M5S 3H8, Canada email: blindert@astro.utoronto.ca, hyee@astro.utoronto.ca\\[\affilskip]

$^2$Carnegie Observatories, 813 Santa Barbara Street, Pasadena, California 91101, USA email:gladders@ociw.edu\\[\affilskip]

$^3$Center for Astrophysics and Space Astronomy, University of Colorado, 389 UCB, Boulder, Colorado 80309-0389, USA email elling@kestrel.colorado.edu\\[\affilskip]
}

\pubyear{2004}
\volume{195}
\pagerange{1--8}
\date{?? and in revised form ??}
\jname{Outskirts of Galaxy Clusters: intense life in the suburbs}
\editors{A. Diaferio, ed.}
\begin{document}

\maketitle

\begin{abstract}
A multi-object spectroscopy follow-up survey of galaxy clusters selected from the Red-sequence Cluster Survey (RCS) is being completed.  About forty clusters were chosen with redshifts from 0.15 to 0.6, and in a wide range of richnesses.  One of the main science drivers of this survey is a study of internal dynamics of clusters. We present some preliminary results for a subset of the clusters, including the correlation of optical richness with mass, and the mass-to-light ratio as a function of cluster mass.
\end{abstract}

\firstsection 

\section{The Survey}
The RCS is a large imaging survey in $R$ and $z^\prime$, which uses galaxy overdensity along with colour information to efficiently detect clusters out to $z\!\!>\!\!1$ (\cite[Gladders \& Yee 2004]{rcs}).  Detection of a cluster with the RCS technique also provides $z_{phot}$, a photometric estimate of the cluster redshift.  In order to study cluster dynamics and galaxy populations, we are completing a spectroscopic follow-up survey of a subset of about $40$ RCS clusters.  These clusters have a wide range of optical richnesses (parametrized by the cluster-centre galaxy correlation amplitude $B_{gc}$; see \cite[Yee \& Ellingson 2003, hereafter Y03]{y03}), and are chosen with redshift estimates $z<0.6$ so as to allow efficient follow-up.  We have performed multi-object spectroscopy of galaxies in the fields of these clusters using the LDSS-2 spectrograph on the Magellan telescopes and the MOS/SIS on the CFHT.  We observe two masks per field, for improved spatial sampling as well as some redundant observations (which help in estimating redshift errors).  In order to obtain spatial coverage out to a few virial radii for each cluster, we observe between one and three fields depending on the redshift and richness of the cluster.  Our selection criteria for mask design depends only on geometry and magnitude: we make no colour cuts, so as to ensure a fair sample for use in studying galaxy populations in these clusters.  Overall, with two to six masks per cluster, roughly 30 spectra per mask with Magellan/LDSS-2 and 100 spectra per mask with CFH/MOS (thanks to a bandlimiting filter which we use to multi-tier our spectra), we have over 7000 spectra in total.  Of course we cannot obtain redshifts from all spectra, nor are all the galaxies cluster members; ultimately we expect to have about 1750 members for the clusters in this survey.

\begin{figure}
\centerline{
\includegraphics[width=0.4\textwidth]{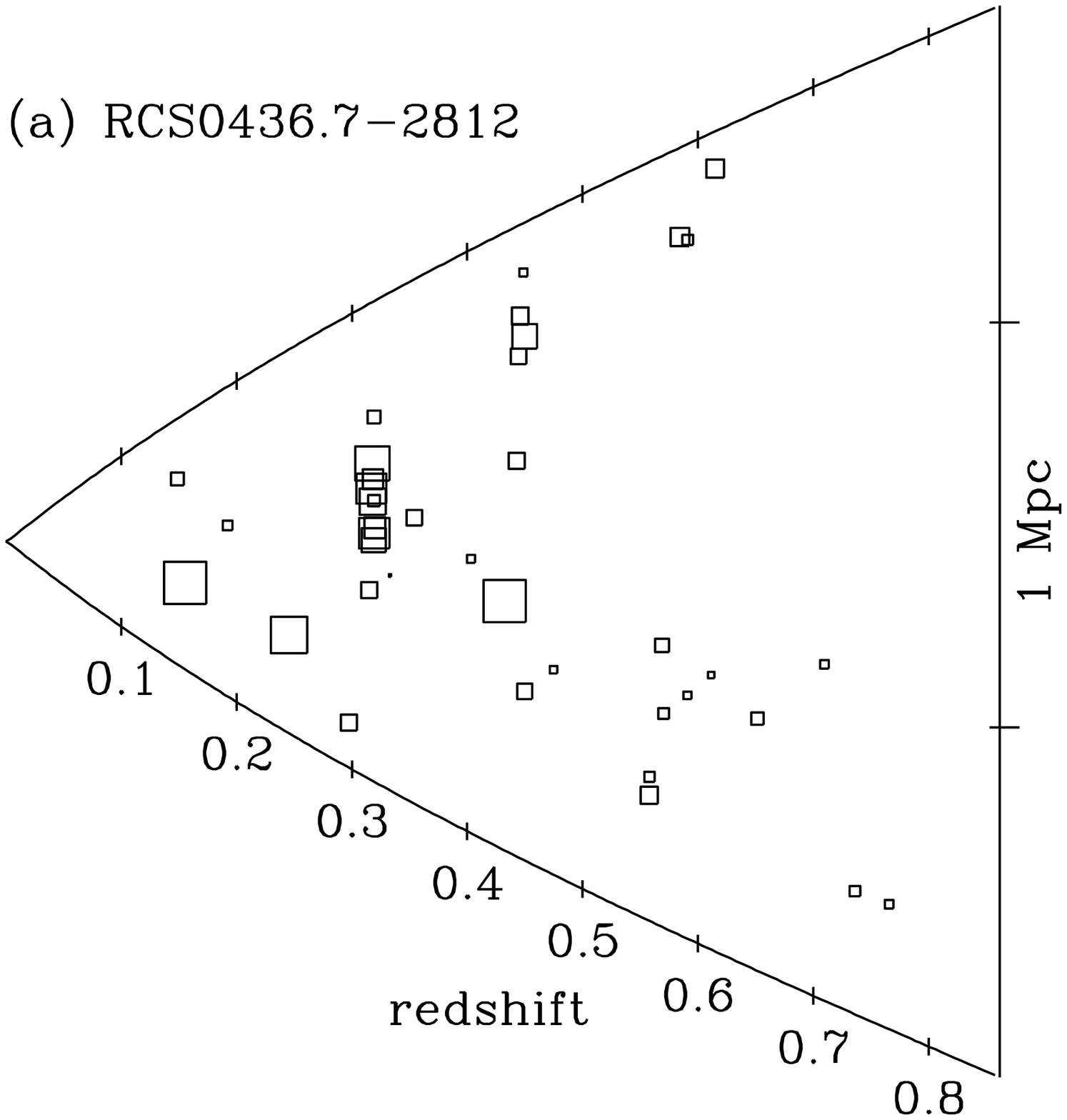}
\includegraphics[width=0.4\textwidth]{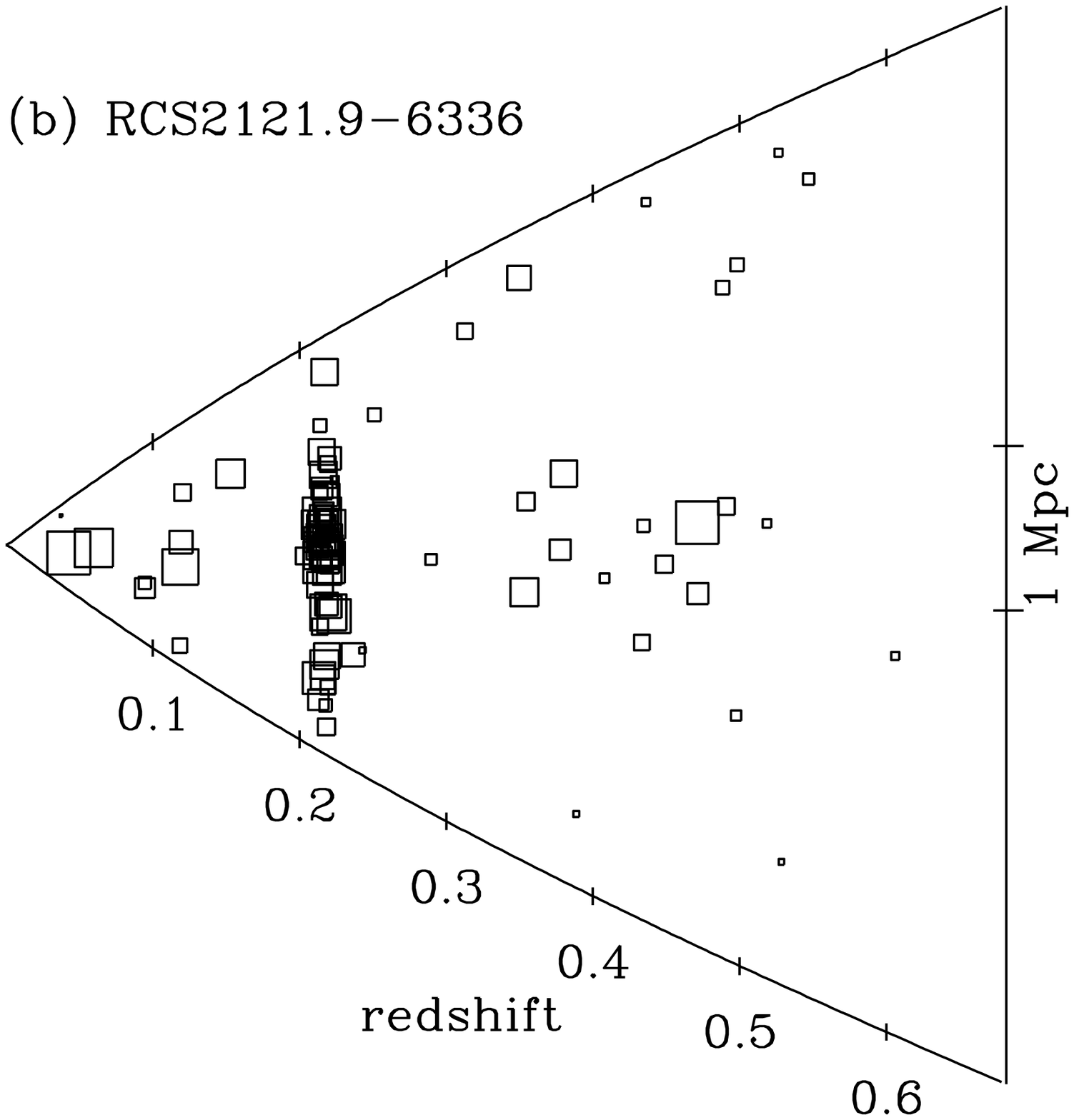}
\\}
\caption{Pie diagrams of two example clusters; diagram (a) shows a poor cluster, while (b) is one of our richest systems. Each square is a galaxy, with size related to apparent magnitude.}
\label{pieplots}
\end{figure}

\section{Preliminary Results}
Here we present first results from 17 of the clusters in our sample for which we have obtained redshifts and computed weights using the CNOC technique (\cite[Yee, Ellingson, \& Carlberg 1996]{cnoc_obs}).  This subset of clusters spans our full redshift and richness ranges: $0.2\!<\!z_{phot}\!<\!0.6$ and $200\!<\!B_{gc}\!<\!1900$.  We use the scheme of \cite{zab} to reject non-members, which is quite strict (and hence may somewhat underestimate cluster properties such as velocity dispersion).  Velocity dispersions are calculated using weights to account for incompleteness (we use magnitude, colour, and geometric weights).  We estimate cluster masses using the formula
$M\!=\!\frac{3}{G}\,{\sigma_1}^2\ r_{vir}$.  The virial radius is usually calculated from pointwise galaxy separations, which can be noisy (\cite[Carlberg et al.~1996, hereafter C96]{c96}); proper calculation of $r_{vir}$ will require further investigation.  In the meantime, we substitute $r_{200}=\frac{\sqrt{3}\sigma_1}{10 H(z)}$ (\cite[Carlberg et al.~1997]{c97}).  K-corrected $R_c$-band cluster luminosities, computed within $r_{200}$, are also estimated using weights.

\begin{figure}
\centerline{
\includegraphics[width=0.6\textwidth, angle=270]{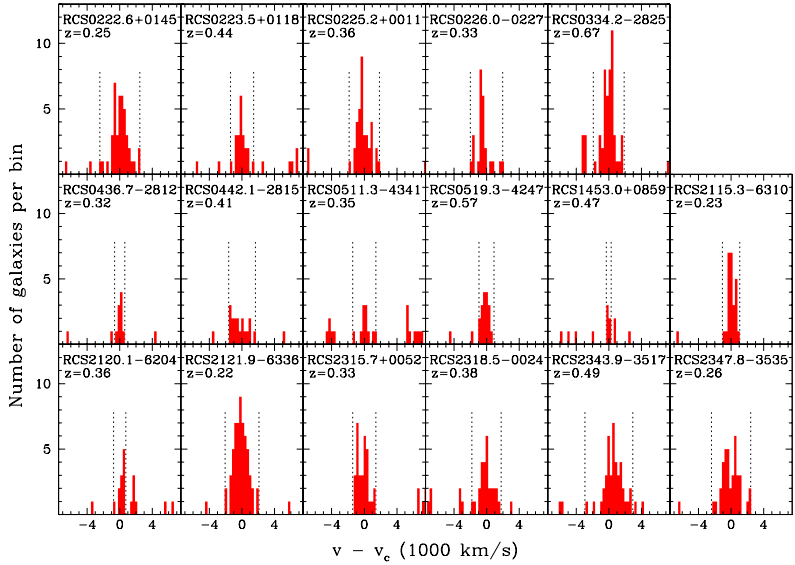}
\\}
\caption{Velocity histograms of the clusters in $300\,{\rm km/s}$ bins.  Dotted lines show the limits chosen by our interloper rejection scheme.}
\label{zhists}
\end{figure}

\subsection{Confirmation of the systems}
Figure~\ref{pieplots} shows pie diagrams for two of the clusters in our subset, one poor and one rich.  Velocity histograms for all 17 clusters are in Figure~\ref{zhists}.  We find many members in all but two systems: RCS0436.7-2812 and RCS1453.0+0859.  The former is a real peak in the galaxy distribution, both on the sky and in velocity space (see Figure~\ref{pieplots}a), but is simply a very poor system: its optical richness is only $B_{gc}\!\!=\!\!256$.  The latter system cannot be confirmed as a physical system: we see no clear peak in either velocity or on the sky.  However, observing conditions were very poor for this cluster and few redshifts were obtained; in fact there does seem to be a small concentration of galaxies on the sky at this location, but redshifts couldn't be determined for any of them.  Thus, though we cannot confirm this cluster candidate, neither can we dismiss it as a `false positive'.

\begin{figure}
\centerline{
\includegraphics[width=0.4\textwidth]{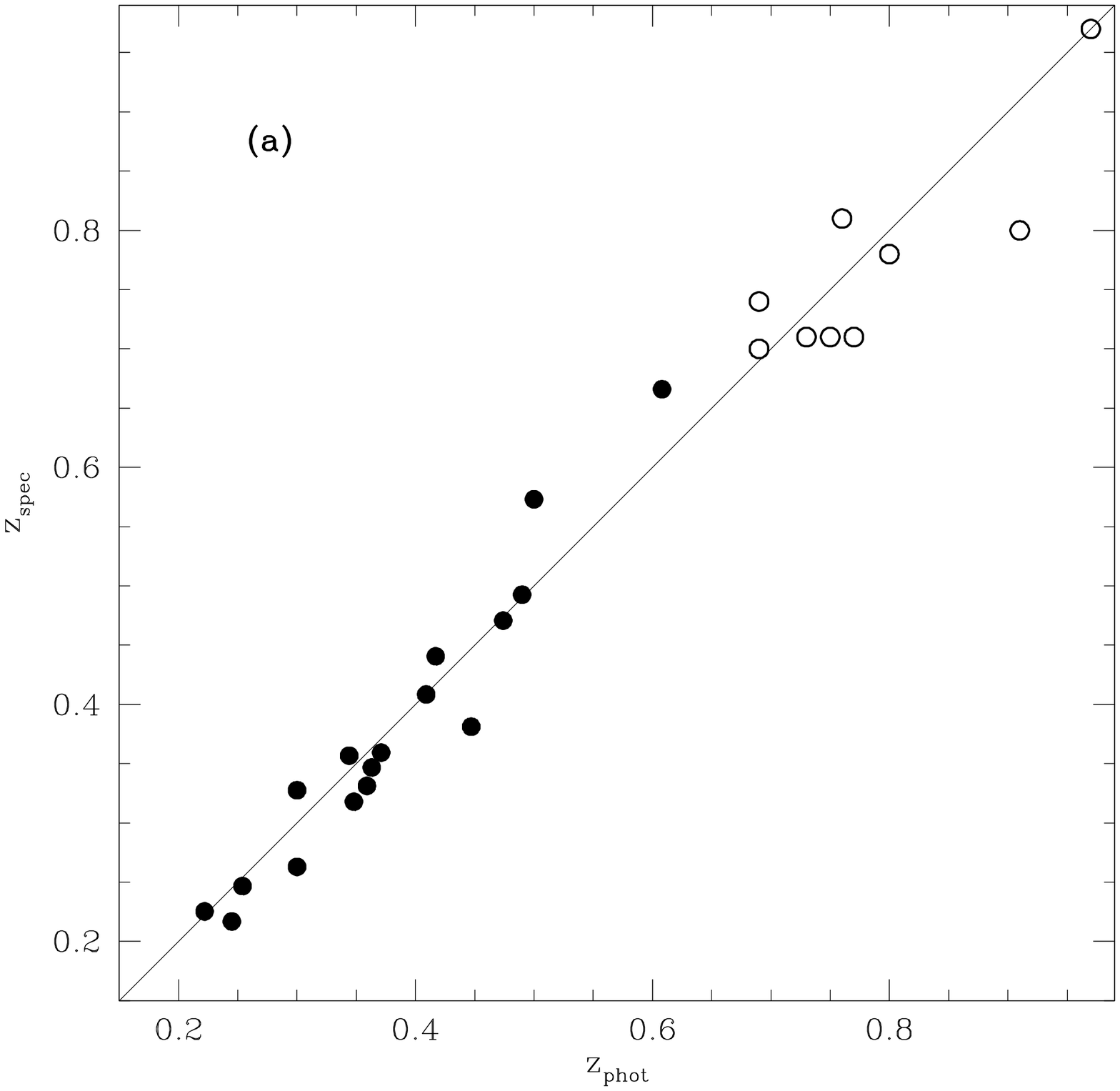}
\includegraphics[width=0.4\textwidth]{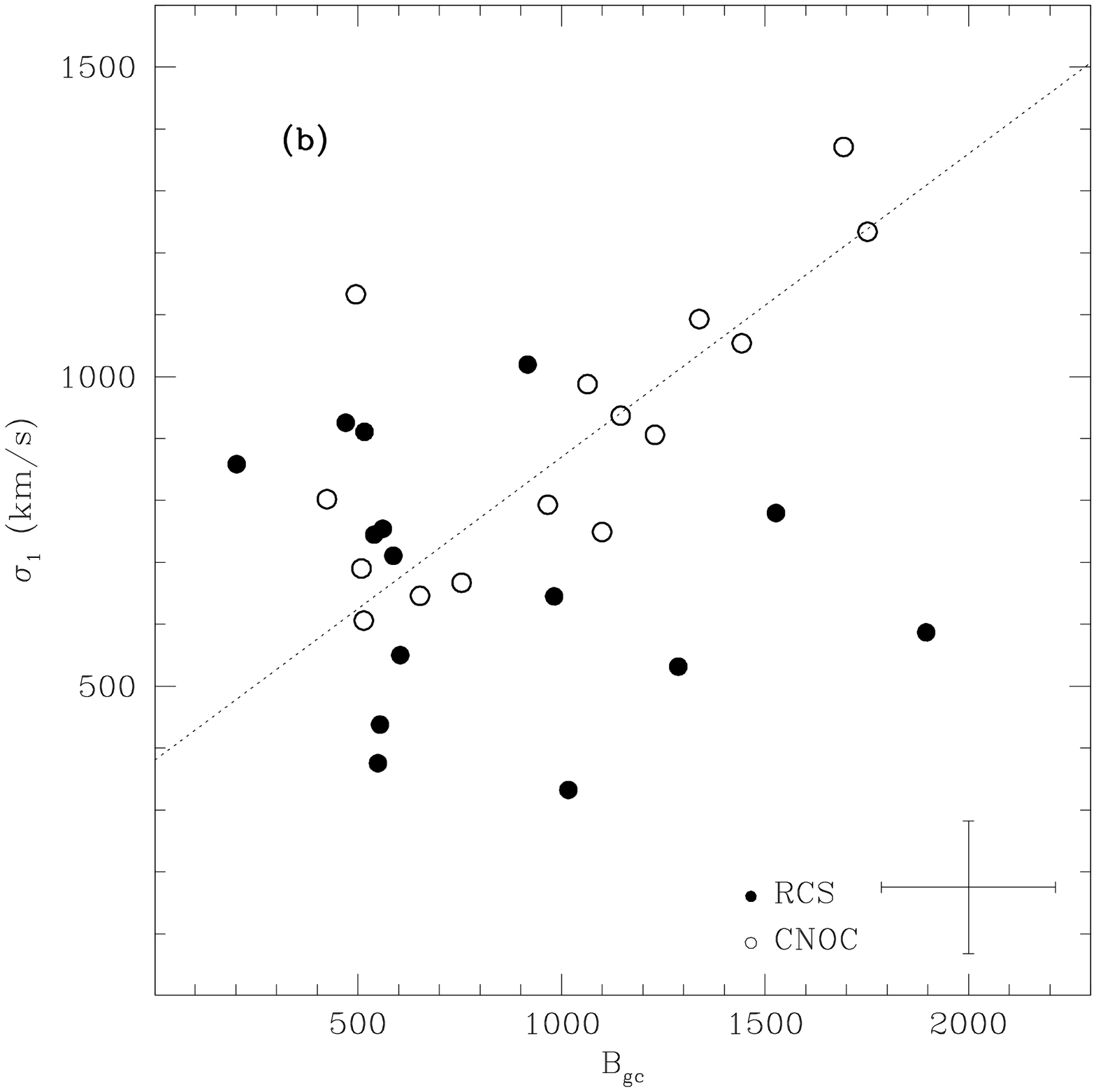}
\\}
\caption{
{\bf (a)} Quality of the RCS redshift estimates. Filled circles are our subsample, and open circles are higher redshift clusters observed with Magellan LDSS-2 by F. Barrientos and M. Gladders.  The solid line is not a fit to the data, but rather the $z_{phot}=z_{spec}$ line.\quad
{\bf (b)} Optical richness versus velocity dispersion.  Filled circles are our subsample, and open circles are CNOC clusters (Y03).  For reference, the best fit to the CNOC clusters is shown as a dotted line.  A typical error bar is shown at bottom right.}
\label{redrich}
\end{figure}

\subsection{Redshifts and richnesses of the clusters}
The RCS technique predicts reliable redshift estimates, even at low redshift where the $R$ and $z^\prime$ filters no longer straddle the $4000\textrm\AA$ break (see Figure~\ref{redrich}a).  The relation between optical richness and velocity dispersion, shown in Figure~\ref{redrich}b, is not as tight.  Though the $B_{gc}$ and $\sigma_1$ values are roughly consistent with those found by Y03, the scatter is larger.  In particular, we find one outlier with very low $\sigma_1$ compared to its optical richness.  It is possible that this is an example for which our interloper rejection algorithm is too strict (see the velocity histogram of RCS0334.2\nolinebreak[4]-\nolinebreak[4]2825 in Figure~\ref{zhists}).  It may be that more sophisticated rejection algorithms will include these galaxies and increase the velocity dispersion of the cluster.

\subsection{Mass-to-light ratios}
\begin{figure}
\centerline{
\includegraphics[width=0.6\textwidth, angle=270]{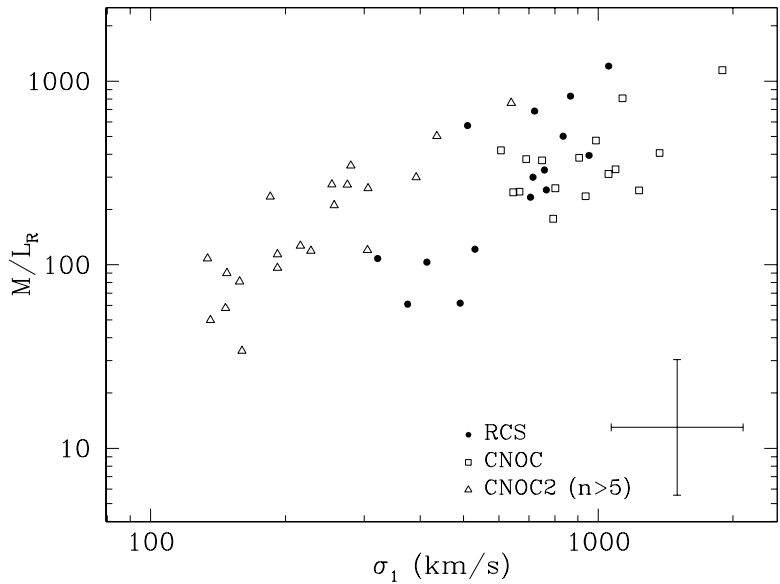}
\\}
\caption{M/L vs. $\sigma_1$.  Filled circles are our clusters; open points are the X-ray selected CNOC clusters (C96) and the groups found by \cite{cnoc2} in the CNOC2 field survey (including only those groups with $N_{mem}\!\!>\!\!5$).  At bottom right we show a typical error bar. Note that errors in the two axes are strongly correlated, since $M$ is computed from $\sigma_1$.}
\label{mlratio}
\end{figure}

The $R$-band mass-to-light ratios are plotted against velocity dispersion in Figure~\ref{mlratio}.  We find an increase in $M/L$ as a function of $\sigma_1$; in other words, more massive clusters have higher mass-to-light ratios.  We caution the reader against over-interpreting this result, since our mass calculations are preliminary and our cluster sample is still relatively small.  However, this result is in qualitative agreement with a comparison of CNOC clusters against CNOC2 groups (also shown in Figure~\ref{mlratio}), as well as with the findings of \cite{eke} on groups in the 2dF galaxy survey and with K-band results for the CAIRNS rich clusters (\cite[Rines et al.~2004, also presented at this conference]{rines}).

\subsection{The ensemble cluster}
\begin{figure}
\centerline{
\includegraphics[width=0.6\textwidth, angle=270]{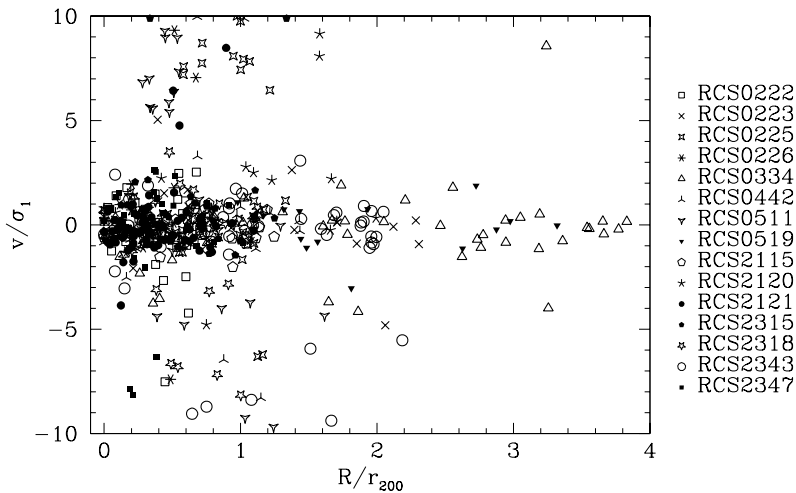}
\\}
\caption{Projected radii and velocities for the ensemble cluster.}
\label{rvplot}
\end{figure}

One of the main goals of this project is to perform detailed dynamical analysis of RCS clusters to compare mass profiles of different cluster masses; since we have relatively few members for any one cluster, it will be necessary to stack our clusters.  We will present detailed analysis of all clusters in a later paper; as an illustration, we here present a single ensemble cluster from the clusters discussed above.  We scale galaxy velocities by $\sigma_1$ and projected cluster-centric radii by $r_{200}$; the result is shown in Figure~\ref{rvplot}.  One can clearly see an overdensity for small clustercentric radii and velocities, within a trumpet-shaped envelope.  This is the redshift-space caustic, which can be used to determine cluster membership as well as the mass profile (\cite[Diaferio \& Geller 1997]{caustic}).

\section{Summary and Future Work}
We are completing a spectroscopic follow-up survey of about 40 RCS clusters in a range of redshifts and richnesses.  We have presented some first results from 
17 of these systems:
\begin{itemize}
\item 16 of the systems are confirmed as real overdensities in redshift space.  Data for the remaining system are too poor to determine whether it is real or not.
\item Predicted redshifts from the RCS method are in good agreement with the spectroscopic redshifts.  Optical richnesses also correlate with velocity dispersions, though weakly.
\item Cluster mass-to-light ratio depends on cluster mass.  This anti-biasing effect is in qualitative agreement with other datasets.
\item Scaled cluster-centric radii and velocities, shown for a single stacked ensemble, show a clear caustic.
\end{itemize}

The analysis we have presented here is quite simple, and several obvious improvements can be made.  Our interloper removal scheme is likely too strict, and we currently use only an estimate of $r_{200}$ in place of the virial radius.  Also, our masses are computed from the virial equation, neglecting the surface term; thus they overestimate the true masses enclosed within $r_{200}$ (C96).  We have 20 more systems to analyse, which will allow us to constrain differences in cluster properties such as mass and mass-to-light profiles with richness.

\end{document}